\documentclass{IJMPB-QST}
\usepackage{graphicx}
\usepackage{amssymb}
\begin{document}

\markboth{Zhe Yang, Ming Gao, Wei Qin.}
{Transfer of high-dimensional quantum state through an $XXZ$-Heisenberg quantum spin chain}

%
\catchline{}{}{}{}{}
%

\title{Transfer of high-dimensional quantum state through an $XXZ$-Heisenberg quantum spin chain}

\author{Zhe Yang}

\address{$^{1}$ State Key Laboratory of Low-Dimensional Quantum Physics and Department of Physics, Tsinghua University, Beijing 100084,
China\\}

\author{Ming Gao}

\address{$^{1}$ State Key Laboratory of Low-Dimensional Quantum Physics and Department of Physics, Tsinghua University, Beijing 100084,
China\\}

\author{Wei Qin}

\address{$^2$ School of Physics, Beijing Institute of Technology, Beijing 100081, China\\
qinwei09@tsinghua.org.cn}

\maketitle


\begin{abstract}
We propose and analyze an efficient high-dimensional quantum state transfer scheme through an $XXZ$-Heisenberg spin chain in an inhomogeneous magnetic field. By the use of a combination of coherent quantum coupling and free spin-wave approximation, pure unitary evolution results in a perfect high-dimensional swap operation between two remote quantum registers mediated by a uniform quantum data bus, and the feasibility is confirmed by numerical simulations. Also, we observe that either the strong $z$-directional coupling or high quantum spin number can partly suppress the thermal excitations and protect quantum information from the thermal noises when the quantum data bus is in the thermal equilibrium state.
\end{abstract}


\section{introduction}
The transfer of quantum state between two distant quantum registers is an essential task of quantum information processing (QIP)\cite{Chuang}. While long-range quantum communication can be realized by the use of photons\cite{photon1,photon2,photon3}, coupled solid-state systems can act as quantum data buses to connect two separated registers for short-range communication, e.g., within a computer. Such data buses have been explored in the context of various quantum systems ranging from trapped ions\cite{ion1,ion2} and super-conducting flux qubits\cite{superconductor1,superconductor2,superconductor3} to cavity arrays\cite{cavity1,cavity2,cavity3} and nanoelectromechanical oscillators\cite{oscillators1}. Due to the ability to provide an alternative to either direct register interactions or an interface between stationary and flying qubits, quantum spin chains have attracted much attention in recent years \cite{qin61,Entang2,Entang3,IJMPB1,IJMPB2,QST1,QST2,QST3,QST4}. In the original scheme\cite{Bose}, S. Bose studied a uniform spin chain of Heisenberg coupling, and quantum information can be efficiently transferred between two ends of the spin channel via natural evolution. Moreover, many strategies aiming to achieve the perfect quantum state transfer (QST) over arbitrary distance have emerged, such as engineering the coupling strength in a way dependent of the chain length\cite{perf11,perf12}, implementing local measurements of individual spins\cite{perf21} and designing some special configurations of spin chains\cite{perf31,perf32,perf33}. Alternatively, coherent quantum coupling has been widely used to achieve high-fidelity QST by tuning the registers to interact weakly with the channel\cite{perf41,perf42,perf43,perf51,perf52}.

Compared to two-dimensional systems working as qubits, high-dimensional systems as qudits also deserve to explore because they can carry large capacity and lead to a further insight into our understanding of quantum physics. Until now, many proposals of quantum computation\cite{CaoY} and quantum communication, e.g. quantum cloning\cite{clon1,clon2}, quantum teleportation\cite{tele1,tele2,tele3}, quantum key distribution\cite{QKDd} and quantum correlation\cite{corr} have been extended to high-dimensional versions. Indeed, with some notable exceptions\cite{high1,high2,high3}, where perfect high-dimensional state transfer over long distance has been implemented by utilizing a repeated measurement procedure or a free spin wave approximation, prior work on perfect QST in coupled-spin systems has primarily focused upon qubits\cite{2d1,2d2,2d3,2d4,2d5,2d6}.

In this paper, we devote our attention to a perfect transfer of high-dimensional quantum state through an $XXZ$-Heisenberg coupling spin chain of arbitrary length in an inhomogeneous magnetic field. On employing the Holstein-Primakoff transformation and the free spin wave approximation, the Hamiltonian takes the form of free bosons and can be diagonalized through an orthogonal transformation. Tuning the register-bus coupling in the $xy$ plane to be much smaller than that within the data bus enables a special data bus collective eigenmode resonating with the two end registers. As a consequence, unitary evolution results in a perfect swap operation between the two registers in the optimal time, and numerical simulations are performed to confirm it. Moreover, we observe that increasing either the strong $z$-directional coupling or high quantum spin number is capable of protecting quantum information from the thermal noises.

The structure of the paper is as follows. In section 2, we introduce the analysis of the model and give the Hamiltonian. In section 3, we show that a high fidelity QST and the thermal effects. Finally, we summarize the whole mechanism and draw our conclusions in the section 4.

\section{Model and Analysis}

\begin{figure}[!ht]
\begin{center}
\includegraphics[width=10cm,angle=0]{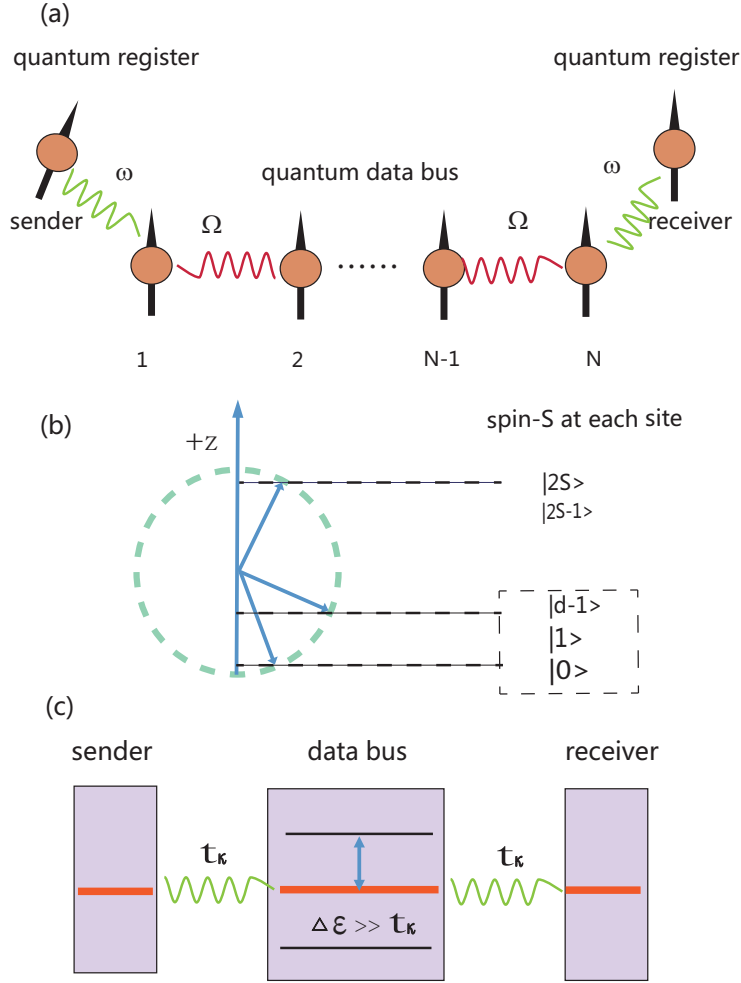}
\caption{(Color online) (a) Shown is a quantum data bus mediating two quantum registers, with an $XXZ$-Heisenberg coupling. We demonstrate that a perfect high-dimensional swap operation between the registers via purely unitary evolution over arbitrary distance by applying an inhomogeneous field. (b) We employ a $d$-dimensional space spanned by the low-lying level states ranging from $|0\rangle$ to $|d-1\rangle$ to encode quantum information as a qudit. The condition $ 2S>>d $ predicts that the spin-wave interaction can be neglected to yield a tight-binding Hamiltonian, which can be diagonalized through an orthogonal transformation. (c) On maintaining $ {\omega _0}/{\Omega _0} << 1$ ,there is a special data bus collective mode being resonantly coupled to the two registers, and off-resonant coupling can be neglected. Therefore, we achieve a high dimensional quantum state transfer protocol through this eigenmode-mediated quantum channel.}\label{f1}
\end{center}
\end{figure}
As shown in Fig. $1(a)$, an $XXZ$-Heisenberg model governs an $(N+2)$-site spin-$S$ chain in an inhomogeneous magnetic field. Only the nearest-neighbor interaction is considered and the system is described by
\begin{equation}\label{H1}
   H = {H_{B}} +{H_I}+ {H_M}.
\end{equation}
The Hamiltonian of the quantum data bus is
\begin{equation}\label{HB}
{H_B} =  - {\Omega _0}\sum\limits_{i = 1}^{N - 1}{(S_i^ + S_{i + 1}^ - + S_i^ - S_{i + 1}^ + )} -{\Omega _z}\sum\limits_{i = 1}^{N - 1} {S_i^z} S_{i + 1}^z,
\end{equation}
where ${\Omega _0}>0$ is the coupling strength in the $xy$-plane and ${\Omega _z}>0$ is that along the $z$-direction. $S_i^\nu$ is the $\nu$  $ (\nu {\rm{ = x,y,z}})$ component of the spin operator $ {{\mathbf{S}}_i} $ at the $i$-th site with $ S_i^ \pm  = S_i^x \pm iS_i^y $.
$H_I$ describes the interaction between the two end registers and the intermediate quantum data bus,
\begin{equation}\label{HI}
{H_I} =  - {\omega _0}(S_s^ + S_1^ -  + S_r^ + S_N^ -  + H.c.) - {\omega _z}(S_s^zS_1^z + S_r^zS_N^z),
\end{equation}
where ${\omega _0}>0$ is the interaction between the sender (receiver) and the quantum data bus in the $xy$-plane and ${\omega _z}>0 $ is that along the $z$-direction.
The Zeeman term reads
\begin{equation}\label{HM}
{H_M} =  - ({B_s}S_s^z + {B_r}S_r^z+\sum\limits_{i = 1}^N {{B_i}S_i^z}),
\end{equation}
with $B_i$ being the local magnetic field on the $ith$-site in the $z$-direction. By implementing the Holstein-Primkoff (HP) transformation $S^{+}_{i}=\sqrt{2S-a_{i}^{\dag}a_{i}}a_{i}$ and $S^{z}_{i}=S-a^{\dag}_{i}a_{i}$, the Hamiltonian can be rewritten in terms of boson operators, and the state of each spin is described by a Fock state instead. In general, the low-lying $d$-dimensional space of the sender is harnessed to encode quantum information, and the input state is $ \left| {{\varphi_s}} \right\rangle  = \sum\nolimits_{u = 0}^{d - 1} {{\alpha _u}{{\left| u \right\rangle }_s}} $, while the data bus and the receiver align in a parallel way being a ferromagnetic order \cite{perf11}, as sketched in Fig. 1(b).

For a spin chain of $N+2$ spins-$S$, the Hilbert space $\mathcal{H }$ is of dimension ${(2S + 1)^{N + 2}}$. The Hamiltonian $H$ preserves the total bosonic number $ N = a_s^\dag {a_s} + a_r^\dag {a_r} + \sum\nolimits_{i = 1}^N {a_i^\dag {a_i}} $ due to $[H,N] = 0$. Therefore,
$\mathcal{H}$ can be decomposed into an invariant subspace $\mathcal{S_G}$ spanned by $|n_{s},n_{1},\cdots,n_{N},n_{r}\rangle$ for $n_{s},n_{i},n_{r}=0,\cdots,d-1$, and the dynamics of the system is completely restricted in the the ${d^{(N + 2)}}$-dimensional subspace $\mathcal{S_G}$.
Suppose that the dimension of the transferred state is much smaller than quantum spin number, i.e., $ d <  < 2S$, the average boson number of each site could be much smaller than $2S$, $\left\langle {a_i^\dag {a_i}} \right\rangle <<2S$. Subsequently, the spin-wave interaction is negligible, such that the HP transformation is simplified to  $S_i^ +  = \sqrt {2S} {a_i}$\cite{HP,HP1}, leading to a bosonized tight-binding Hamiltonian
\begin{equation}\label{HF}
\begin{array}{l}
{H_B} =  - 2{\Omega _0}S\;\sum\nolimits_{i = 1}^{N - 1} {({a_i}^{\dag}a_{i + 1} + H.c.)}
- {\Omega _z}\sum\nolimits_{i = 1}^{N - 1} {[{S^2} - S(a_i^\dag {a_i} + a_{i + 1}^\dag {a_{i + 1}})]} ,\\
\\
{H_I} =  - 2{\omega _0}S(a_s^\dag {a_1} + a_r^\dag {a_N} + H.c.) - {\omega _z}[2{S^2} - S(a_s^\dag {a_s} + a_1^\dag {a_1} + a_r^\dag {a_r} + a_N^\dag {a_N})],\\
\\
{H_M} =  - \left[B_{s}\left(S-a_{s}^{\dag}a_{s}\right)+B_{r}\left(S-a_{r}^{\dag}a_{r}\right)+\sum_{i=1}^{N} {{B_i}(S - a_i^\dag {a_i})}\right].
\end{array}
\end{equation}
In order to achieve an efficient high-dimensional state transfer, we choose
\begin{equation}\label{B}
\begin{array}{l}
{B_s} = {B_r} = 2{\Omega _z}S, \\
{B_1} = {B_N} = {\Omega _z}S, \\
{B_2} =\cdots={B_{N - 1}} = {\omega _z}S,
\end{array}
\end{equation}
and apply the following orthogonal transformation\cite{Chuang,perf51,XXd}
\begin{equation}\label{tran}
a_i^\dag  = \sqrt {\frac{2}{{N + 1}}} \sum\limits_{k = 1}^N {\sin } \frac{{ik\pi }}{{N + 1}}c_k^\dag, \quad i=1,...,N,
\end{equation}
the Hamiltonian $H$ is transformed to
\begin{equation}\label{HHH}
\begin{array}{l}
H{\rm{ = }}\sum\limits_{k = 1}^N {[({\varepsilon _k} + \Gamma )c_k^\dag {c_k}]} {\rm{ + }}\Gamma (a_s^\dag {a_s} + a_r^\dag {a_r})
{\rm{ + }}\sum\limits_{k = 1}^N {{t_k}[a_s^\dag {c_k} + {{\left( { - 1} \right)}^{k - 1}}a_r^\dag {c_k} + H.c]}.
\end{array}
\end{equation}
where ${\varepsilon _k} =  - 4{\Omega _0}S\cos (\frac{{k\pi }}{{N + 1}})$, $ {t_k} =- 2\omega_{0}S\sqrt{\frac{2}{N+1}}\sin (\frac{{k\pi }}{{N + 1}})$ and $\Gamma  = (2{\Omega _z} + {\omega _z})S$. Note that the choice of the nonuniform field is applicable for only $N\geq3$, and in the special case of $N=1$, the field can be chosen as $B_{s}=B_{r}=\omega_{z}S+h$ and $B_{1}=h$ with $h\geq0$.

\section{Quantum state transfer and the thermal effects}
By restricting our discussion to a case of odd $N$ chains, there exists a zero-energy data bus collective mode, corresponding to $\kappa {\rm{ = (N + 1)/2}}$, being resonantly coupled to the two end registers with strength ${t_\kappa } =  - 2{\omega _0}S/A $ where $A = \sqrt {{{(N + 1)}\mathord{\left/{\vphantom {{(N + 1)} 2}}\right.\kern-\nulldelimiterspace}2}}$. Under the assumption that $ {\omega _0}/{\Omega _0} <  < 1/\sqrt N $, off-resonant coupling can be neglected as a result of ${t_\kappa } \ll |{\varepsilon _\kappa } - {\varepsilon _{\kappa  \pm 1}}|$, such that evolution dynamics behaves as an effective model in which only the two end registers and the $\kappa$-th collective mode are involved, as illustrated schematically in Fig. 1(c). In this case, the effective Hamiltonian
\begin{equation}\label{Heff}
\begin{array}{l}
{H_{\text{eff}}}{\rm{ = }}\Gamma (c_\kappa ^\dag {c_\kappa } + a_s^\dag {a_s} + a_r^\dag {a_r})+{t_\kappa}[a_s^\dag {c_\kappa} + {\left( { - 1} \right)^{\kappa - 1}}a_r^\dag {c_\kappa} + H.c]
\end{array}
\end{equation}
governes the evolution of the system. In the Heisenberg picture, the operators should evolve in the full space associated with $H_{\text{eff}}$.  Thus by choosing evolution time $\tau  \equiv \pi /\sqrt 2 {t_\kappa }$, it yields
\begin{equation}\label{as}
\begin{array}{l}
a_s^\dag ({\tau}) = {( - 1)^\kappa }{e^{ - i\Gamma \tau }}a_r^\dag ,\;\;\;a_r^\dag ({\tau}) = {( - 1)^\kappa }{e^{ - i\Gamma \tau }}a_s^\dag,
\end{array}
\end{equation}
which reveals that the quantum state of the sender can be perfectly transfered to the receiver in the optimal time $\tau $. A swap gate has been established between the sender and the receiver, up to an additional phase independent of the sent state. Without decoherence, our scheme can, in principle, achieve perfect QST in spin chains of arbitrary length. However, the optimal time should be much shorter than the coherence time when decoherence is present, and the chain length is therefore limited.

To confirm the efficiency of our method, numerical simulations are performed. Initially, the whole system, including the two end registers and the intermediate data bus, is in a product state
\begin{equation}\label{State}
|\psi\left(0\right)\rangle =|\varphi\rangle_{s}|0\rangle_{\text{bus}}^{\otimes N}|0\rangle_{r},
\end{equation}
and $|0\rangle_{\text{bus}}^{\otimes N}=|0\rangle_{1}\otimes\cdots\otimes |0\rangle_{N}$. In general, the state of the receiver at time $ t $ is a mixed state ${\rho _r}(t )$, which can be obtained by tracing off the other sites ${\rho _r}(\tau ) = \text{Tr}{_{\hat r}}({e^{ - iHt}}\left| {\psi (0)} \right\rangle \langle \psi (0)|{e^{iHt }})$.
\begin{figure}[!ht]
\begin{center}
\includegraphics[width=0.6\textwidth,angle=0]{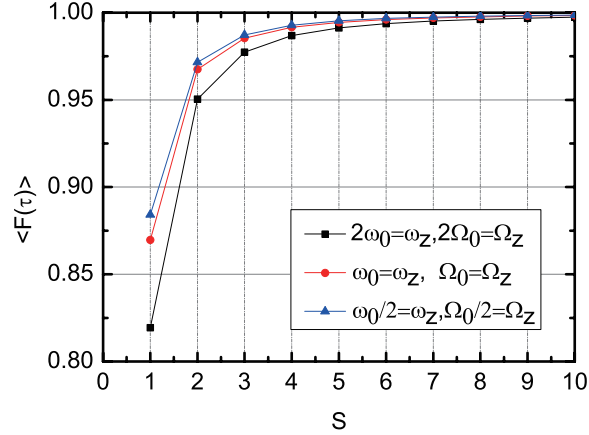}
\caption{(Color online) The average fidelity varies as a function of quantum spin number $S$ with $N=3$, $d=3$ and ${\omega _0}/{\Omega _0} = 0.1$ for three values of the $z$-directional coupling. Here, the evolution time is the optimal time $\tau$.}\label{f2}
\end{center}
\end{figure}

A complex projective space $\mathbb{C}{P^{d - 1}} $ is constructed by the set of the pure states of a $d$-dimensional Hilbert space. According to Hurwitz parametrization method, a pure $d$-dimensional state must be described by $2(d-1)$ parameters including $d-1$ azimuthal angles ${\theta _i}$  and $d-1$ polar angles $ {\varphi _i}$
\begin{equation}\label{qudit}
\left| \Psi  \right\rangle  = (\cos {\theta _{d - 1}},\sin {\theta _{d - 1}}\cos {\theta _{d - 2}}{e^{i{\varphi _{d - 1}}}}, \sin {\theta _{d - 1}}\sin {\theta _{d - 2}}\cos {\theta _{d - 3}}{e^{i{\varphi _{d - 2}}}},...,\prod\limits_{i = 1}^{d - 1} {\sin {\theta _i}} {e^{i{\varphi _1}}})
\end{equation}
with ${\theta _i} \in [0,\frac{\pi }{2}]$ and $ {\varphi _i} \in [0,2\pi )$.
The fidelity between the sent state of the sender and the received state of the receiver at time $\tau$ is given by $F(\tau ){ = _s}\langle \varphi |{\rho _r}(\tau ){\left| \varphi  \right\rangle _s}$. Correspondingly, the average fidelity over all possible input pure states is
\begin{equation}\label{Faver}
\left\langle {F(\tau )} \right\rangle  = \frac{1}{V}\int_{} {F(\tau )} dV.
\end{equation}
Here, $V=\pi^{d-1}/\left(d-1\right)$ is the total volume of the manifold of pure states, and $ dV =\prod_{p=1}^{d-1}\cos\vartheta_{p}\left(\sin\vartheta\right)^{2p-1}d\vartheta_{p}d\varphi_{p}$ is the volume element\cite{highst}. In fact, the average fidelity is a generalization of the usual Bose formula\cite{Bose}, i.e. in the case of $d=2$, Eq. (\ref{Faver}) takes the same form as the average fidelity of the case of qubit.

In Fig. 2 the average fidelity varies as a function of quantum spin number $S$ when $N=3$ and $d=3$ with ${\omega _0}/{\Omega _0} = 0.1$. The numerical results are based on the Hamiltonian of Eq. (1), and three different $z$-directional coupling strengths are chosen to demonstrate the feasibility of the method. We observe that the average fidelity increases with $S$, and when ${\omega _0}/{\Omega _0} <<1$ and $d<<2S$, the average fidelity nearly tends to one, e.g.,
in a case of $S=10$, $\left\langle {F(\tau )} \right\rangle$ is $0.9974$ (black line), $0.9984$ (red line), and $0.9986$ (blue line). The leakage of quantum information results from either the off-resonant coupling or the spin-wave interaction.

In the following, an investigation on the thermal effects will be numerically given when the quantum data bus is in a thermal equilibrium state described by
\begin{equation}\label{pb}
{\rho _B} = \frac{1}{Z}e^{-H_{B}/T},
\end{equation}
where $Z = \textnormal{tr}({e^{ - {H_B}/T}})$ characterizes a partition function and $T$ represents the temperature.
\begin{figure}[!ht]
\begin{center}
\includegraphics[width=0.55\textwidth,angle=0]{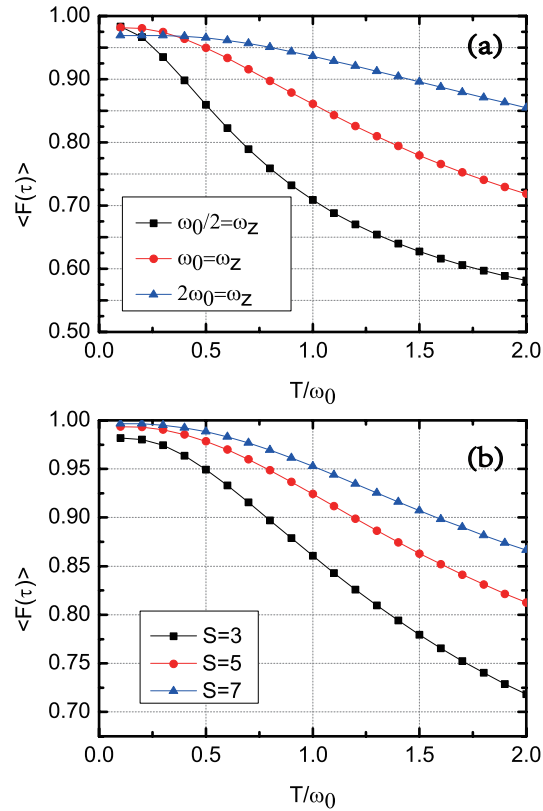}
\caption{(Color online) The average fidelity varies as a function of temperature with $N=1$ and $d=3$ for either (a) three $z$-directional coupling strengths in a case of $S=3$, or (b) three quantum spin numbers in a case of ${\omega _0} = {\omega _z}$. Here, $h=\omega_{z}S$ and the evolution time is the optimal time $\tau$.}\label{f4}
\end{center}
\end{figure}
The density matrix of the whole system is in a product state,
\begin{equation}\label{p0}
\rho (0) = \sum\limits_{\mu ',\mu  = 0}^{d - 1} {{\alpha _\mu }\alpha _{\mu '}^*{{\left| \mu  \right\rangle }_s}{{\left| 0 \right\rangle }_r}\langle \mu '{|_s}\langle 0{|_r}}  \otimes {\rho _B}.
\end{equation}

In Fig. \ref{f4} we plot the average fidelity as a function of the temperature for a bus of length $N=1$ initially in its thermal equilibrium state: $F\langle\tau\rangle$ decreases with $T/\omega_{0}$ owing to the validity of free spin wave approximation only in the low boson excitation regime, however, increasing either $\omega_{z}$ or $S$ can depress the thermal noises to prevent quantum information from leaking. In Fig. \ref{f4}(a), it should be noted that the $z$-directional coupling contains the spin-wave interaction found in nonlinear terms of the HP transformation, and such coupling can lead to the leakage of quantum information, specially at very low temperature range. However, with the temperature increasing, the $z$-directional coupling can effectively cope with the thermal effects, and provide the protection for quantum information instead. Moreover, from the model featured by $H$ of Eq. (\ref{HHH}), both the $z$-directional coupling and the magnetic field result in $\Gamma$ being capable of protecting quantum information, in similar to the magnetic field applied on an $XX$ coupling spin chain.

\section{Summary}
In this paper a quantum state transfer protocol has been studied through an $XXZ$ coupling spin chain in the presence of an inhomogeneous magnetic field. Upon harnessing coherent quantum coupling and free spin-wave approximation, off-resonant couplings and spin-wave interactions can be ignored, and consequently, an arbitrary unknown high-dimensional quantum state can be transferred between two remote registers with high fidelity via purely dynamical evolution. When the quantum data bus is in the thermal equilibrium state, the effects of the temperature on the state transfer protocol have also been numerically studied. In contrast to previous work on $XX$ coupling spin chains, an additional $z$-directional coupling can depress the thermal excitations and partly counteract the thermal effects to ensure the feasibility of the present method. With its scalability and robustness, this protocol may be applicable in a high-dimensional solid device for quantum information processing.


\section*{Acknowledgments}

\quad We gratefully thank Chao Lian, Shuzhe Shi and Hui Li for helpful discussions.


\section*{References}
\begin{thebibliography}{99}

\bibitem{Chuang} M. A. Nielsen and I. L. Chuang, \emph{Quantum Computation and Quantum Information} (UK: Cambridge University Press, 2000).
\bibitem{photon1} K. Mattle, H. Weinfurter, P. G. Kwiat, and A. Zeilinger, Phys. Rev. Lett. {\bf 76}, 4656 (1996).
\bibitem{photon2} J. I. Cirac, P. Zoller, H. J. Kimble, and H. Mabuchi, Phys. Rev. Lett.  {\bf 78}, 3221 (1997).
\bibitem{photon3} T. Jennewein, C. Simon, G. Weihs, H. Weinfurter and A. Zeilinger, Phys. Rev. Lett. {\bf 84}, 4729 (2000).
\bibitem{ion1} D. Kielpinski, C. Monroe, and D. J. Wineland, Nature (London) {\bf 417}, 709 (2002).
\bibitem{ion2} F. Schmidt-Kaler, H. Haffner, M. Riebe, S. Gulde, G. P. T. Lancaster, T. Deuschle, C. Becher, C. F. Roos, J. Eschner, and
R. Blatt, Nature (London) {\bf 422}, 408 (2003).
\bibitem{superconductor1} M. A. Sillanp\"{a}\"{a}, J. I. Park, and R. W. Simmonds, Nature (London) {\bf 449}, 438 (2007).
\bibitem{superconductor2} J. Q. You and F. Nori Nature (London) {\bf 474}, 589 (2011).
\bibitem{superconductor3} J. Majer, J. M. Chow, J. M. Gambetta, Jens Koch, B. R. Johnson, J. A. Schreier, L. Frunzio, D. I. Schuster, A. A. Houck, A. Wallraff, A. Blais, M. H. Devoret, S. M. Girvin, and R. J. Schoelkopf, Nature (London) {\bf 449}, 443 (2007).
\bibitem{cavity1} C. D. Ogden, E. K.Irish, and M. S. Kim, Phys. Rev. A {\bf 78}, 063805 (2008).
\bibitem{cavity2} G. D. de Moraes Neto, M. A. de Ponte, and M. H. Y. Moussa, Phys. Rev. A {\bf 84}, 032339 (2011).
\bibitem{cavity3} Y. Liu, and D. L. Zhou, New J. Phys. 17(1), 013032 (2015).
\bibitem{oscillators1} J. Eisert, M. B. Plenio, S. Bose, and J. Hartley, Phys. Rev. Lett. {\bf 93}, 190402 (2004).
\bibitem{Bose} S. Bose, Phys. Rev. Lett. {\bf 91}, 207901, (2003).
\bibitem{perf11} M. Christandl, N. Datta, A. Ekert, and A. J. Landahl, Phys. Rev. Lett. {\bf 92}, 187902 (2004).
\bibitem{perf12} M. Christandl, N. Datta, T. C. Dorlas, A. Ekert, A. Kay, and A. J. Landahl, Phys. Rev. A {\bf 71}, 032312 (2005).
\bibitem{perf21} F. Verstraete, M. A. Mart\'{i}n-Delgado, and J. I. Cirac, Phys. Rev. Lett. {\bf 92}, 087201 (2004).
\bibitem{perf31} D. Burgarth and S. Bose, Phys. Rev. A {\bf 71}, 052315 (2005).
\bibitem{perf32} Y. Li, T. Shi, B. Chen, Z. Song, and C. P.  Sun, Phys. Rev. A {\bf 71}, 022301 (2005).
\bibitem{perf33} T. J. Osborne and N. Linden, Phys. Rev. A {\bf 69}, 052315 (2004).
\bibitem{perf41} A. W\'{a}jcik, T. {\L}uczak, P. Kurzy\'{n}ski, A. Grudka, T. Gdala, and M. Bednarska, Phys. Rev. A {\bf 72}, 034303 (2005).
\bibitem{perf42} A. W\'{a}jcik, T. {\L}uczak, P. Kurzy\'{n}ski, A. Grudka, T. Gdala, and M. Bednarska, Phys. Rev. A {\bf 75}, 022330 (2007).
\bibitem{perf43} L. Campos Venuti, C. Degli Esposti Boschi, and M. Roncaglia, Phys. Rev. Lett. {\bf 99}, 060401 (2007).
\bibitem{perf44} L. Campos Venuti, S. M. Giampaolo, F. Illuminati, and P. Zanardi, Phys. Rev. A {\bf 76}, 052328 (2007).
\bibitem{perf51} N. Y. Yao, L. Jiang, A. V. Gorshkov, Z.-X. Gong, A. Zhai, L.-M. Duan, and M. D. Lukin, Phys. Rev. Lett. {\bf 106}, 040505 (2011).
\bibitem{perf52} N. Y. Yao, Z.-X. Gong, C. R. Laumann, S. D. Bennett, and L.-M. Duan, and M. D. Lukin, and L. Jiang, and A. V. Gorshkov, Phys. Rev. A {\bf 87}, 022306 (2013).
\bibitem{qin61} W. Qin, C. Wang, Y. Cao, G. L. Long, Phys. Rev. A {\bf 89}, 062314 (2014).
\bibitem{Entang2} L. Campos Venuti, S. M. Giampaolo, F. Illuminati, and P. Zanardi, Phys. Rev. A. {\bf 76}, 052328 (2007).
\bibitem{Entang3} R. H. Crooks and D. V. Khveshchenko, Phys. Rev. A {\bf 77}, 062305 (2008).
\bibitem{IJMPB1} T. J. G. Apollaro, S. Lorenzo, and F. Plastina, Int. J. Mod. Phys. B {\bf 27}, 1345035 (2013).
\bibitem{IJMPB2} J. Liu, G. F. Zhang, and Z. Y. Chen, Int. J. Mod. Phys. B {\bf 24}, 1279 (2010).
\bibitem{QST1} S. Paganelli, F. De Pasquale and G. L. Giorgi, Phys. Rev. A {\bf 74}, 012316 (2006).
\bibitem{QST2} S. Lorenzo, T. J. G. Apollaro, A. Sindona and F. Plastina, Phys. Rev. A {\bf 87}, 042313 (2013).
\bibitem{QST3} S. Lorenzo, T. J. G. Apollaro, S. Paganelli, G. M. Palma and F. Plastina, Phys. Rev. A {\bf 91}, 042321 (2015).
\bibitem{QST4} S. J. Large, M. S. Underwood and D. L. Feder, Phys. Rev. A {\bf 91} 032319, (2015).
\bibitem{CaoY} Y. Cao, S. G. Peng, C. Zheng, and G. L. Long, Commun. Theor. Phys. {\bf 55}, 790 (2011).
\bibitem{clon1} R. F. Werner, Phys. Rev. A {\bf 58}, 1827 (1998).
\bibitem{clon2} M. Keyl and R. F. Werner, J. Math. Phys. {\bf 40}, 3283 (1999).
\bibitem{cg} A. Acin, N. Gisin and V. Scarani, Quantum Inf. Comput. {\bf 3}, 563 (2003).
\bibitem{tele1} G. Rigolin, Phys. Rev. A {\bf 71}, 032303 (2005).
\bibitem{tele2} X. Ge and Y. Shen, Phys. Lett. B {\bf 606}, 184 (2005).
\bibitem{tele3} M. Jiang, X. Huang, L. L. Zhou, Y. M. Zhou, and J. Zeng, Chin. Sci. Bull. {\bf 57}, 2247 (2012).
\bibitem{QKDd} V. Karimipour, A. Bahraminasab, and S. Bagherinezhad, Phys. Rev. A {\bf 65}, 052331 (2002).
\bibitem{corr} H. Li, Y. S. Li, S. H. Wang, and G. L. Long, Commun. Theor. Phys. {\bf 61}, 273 (2014).
\bibitem{high1} A. Bayat, Phys. Rev. A {\bf 89}, 062302 (2014).
\bibitem{high2} W. Qin, C. Wang, and G. L. Long, Phys. Rev. A {\bf 87}, 012339 (2013).
\bibitem{high3} W. Qin, J. L. Li, and G. L. Long, Chin. Phys. B {\bf 24}, 040305 (2015).
\bibitem{2d1} T. J. G. Apollaro, L. Banchi, A. Cuccoli, R. Vaia and P. Verrucchi, Phys. Rev. A {\bf 85}, 052319 (2012).
\bibitem{2d2} L. Banchi, T. J. G. Apollaro, A. Cuccoli, R. Vaia, and P. Verrucchi, New J. Phys. {\bf 13}, 123006 (2011).
\bibitem{2d3} K. Korzekwa, P. Machnikowski and P. Horodecki, Phys. Rev. A, {\bf 89}, 062301 (2014).
\bibitem{2d4} Z. C. Shi, X. L. Zhao, and X. X. Yi, Phys. Rev. A, {\bf 91}, 032301 (2015).
\bibitem{2d5} S. Paganelli, S. Lorenzo, T. J. Apollaro, F. Plastina and G. L. Giorgi, Phys. Rev. A, {\bf 87}, 062309 (2013).
\bibitem{2d6} W. Qin, C. Wang and X. Zhang, Phys. Rev. A, {\bf 91}, 042303 (2015).
\bibitem{HP} T. Holstein and H. Primakoff, Phys. Rev. {\bf 58}, 1098 (1940).
\bibitem{HP1} J. M. Ziman {\it Principles of the theory of solids, second edition}
(Cambridge University Press, Cambridge, UK, 1972).
\bibitem{XXd} E. Lieb, T. Schultz, and D. Mattis, Ann. Phys. (NY)  {\bf 16}, 407 (1961).
\bibitem{highst} K. \.{Z}yczkowski and H. Sommers, J. Phys. A  {\bf 34}, 7111 (2001).
\end{thebibliography}
\end{document}